\documentclass[12pt]{article}
\def\slash#1{\setbox0=\hbox{$#1$}#1\hskip-\wd0\hbox to\wd0{\hss\sl/\/\hss}}
\usepackage{epsf, cite, amsmath, amssymb}
\usepackage{epsfig}
\usepackage[all]{xy}
\usepackage{subfigure}
\usepackage{graphicx}

\makeatletter
\renewcommand\section{\@startsection {section}{1}{\z@}%
                                   {-3.5ex \@plus -1ex \@minus -.2ex}
                                    {2.3ex \@plus.2ex}%
                                   {\normalfont\large\bfseries}}
\renewcommand\subsection{\@startsection{subsection}{2}{\z@}%
                                     {-3.25ex\@plus -1ex \@minus -.2ex}%
                                     {1.5ex \@plus .2ex}%
                                     {\normalfont\bfseries}}
\makeatother

\let\non\nonumber

\newcommand{\bea}{\begin{eqnarray}}
\newcommand{\eea}{\end{eqnarray}}
\newcommand{\be}{\begin{equation}}
\newcommand{\ee}{\end{equation}}

\newcommand{\p}{\partial}

\newcommand{\s}{\sigma}

\newcommand{\C}[1]{$(\ref{#1})$}

\begin{document}

\begin{titlepage}

\begin{center}



\vskip 2 cm
{\Large \bf The $D^{10} {\cal{R}}^4$ term in type IIB string theory}\\
\vskip 1.25 cm { Anirban Basu$^{a}$\footnote{email: abasu@ias.edu}
}\\
{\vskip 0.75cm
$^a$ Institute for Advanced Study, Princeton, NJ 08540, USA\\
}

\end{center}

\vskip 2 cm

\begin{abstract}
\baselineskip=18pt

The modular invariant coefficient of the $D^{2k} {\cal{R}}^4$ term 
in the effective action of type IIB superstring theory is expected to 
satisfy Poisson equation on the fundamental 
domain of $SL(2,\mathbb{Z})$. Under certain assumptions, we obtain the 
equation satisfied by $D^{10} {\cal{R}}^4$
using the tree level and one loop results for four graviton scattering in type II string 
theory. This leads to the conclusion that the perturbative contributions to $D^{10} {\cal{R}}^4$ 
vanish above three loops, and also predicts the coefficients at two and three loops.

\end{abstract}

\end{titlepage}

\pagestyle{plain}
\baselineskip=18pt

\section{Introduction}

Understanding higher derivative corrections to the supergravity action is an important
problem in string theory as well as in M theory. In particular, the large amount of 
supersymmetry and the exact $SL (2, \mathbb{Z})$ invariance of type IIB superstring theory
allows us to study in detail certain higher derivative corrections to the type IIB 
supergravity action. Considering
configurations where the axion--dilaton is constant, the effective action 
of type IIB superstring theory can be schematically written as

\be \label{action}
S = \frac{S^{(0)}}{\alpha'^4}  + \frac{S^{(3)}}{\alpha'}  + \sum_{n=1}^{\infty}
\alpha'^n S^{(n+4)} + \dots, \ee  
where $\ldots$ represents the terms that are non--perturbative in $\alpha'$. 
In the expression in \C{action}, $S^{(0)}$ is
the type IIB supergravity action, and the first corrections to it are at 
$O(1/\alpha')$. (Unlike the usual treatments, if the 
axion--dilaton is not constant, the structure of the terms in the effective action 
is different~\cite{Greentalk}.) There are 
certain terms in the effective action that are tractable because they satisfy various conjectured 
(which have been proven in some cases) non--renormalization theorems. 
In fact, the axion--dilaton dependence of these terms can be completely determined in some cases, 
as we briefly review below (see~\cite{Green:1999qt} for various details). In the discussion below, we 
shall denote the type IIB axion--dilaton by the complexified coupling 
\be \tau \equiv \tau_1 + i \tau_2 = C^{0} + \frac{i}{e^{\phi}},\ee
where $\phi$ is the dilaton and $C^0$ is the Ramond--Ramond pseudoscalar field. 
At $O(1/\alpha')$, one of the protected terms in the effective action is given in the string 
frame by\footnote{Note that ${\cal{R}}^4$ is modular invariant only after transforming to the
Einstein frame.}
\be \label{susyone}
\frac{1}{\alpha'} \int d^{10} x \sqrt{-g} e^{-\phi/2}Z_{3/2} (\tau, \bar\tau) {\cal{R}}^4, \ee
where ${\cal{R}}^4$ involves four powers of the Weyl curvature tensor. Its coupling dependence is 
given by the non--holomorphic
Eisenstein series of modular weight (0,0)
\be Z_s (\tau,\bar\tau) = \sum_{(m,n) \neq (0,0)}
\frac{\tau_2^s}{\vert m + n\tau\vert^{2s}},\ee  
for the value $s=3/2$. Expanding $Z_{3/2}$ at weak coupling, one can show that
it receives only two perturbative contributions at tree level and at one loop, 
as well as an infinite 
number of non--perturbative contributions due to 
D--instantons~\cite{Green:1997tv,Green:1997as,Green:1998by,D'Hoker:2005jc,Iengo:2002pr,Berkovits:2004px}. 
This kind of dramatic non--renormalization is a generic 
property which characterizes these protected terms. There are other terms 
at $O(1/\alpha')$ in the effective action which are related to 
\C{susyone} by supersymmetry, which also satisfy non--renormalization theorems. For example,
one such term is given by
\be 
\frac{1}{\alpha'} \int d^{10} x \sqrt{-g} e^{-\phi/2}  f^{(12,-12)}(\tau, \bar\tau) 
\lambda^{16}, \ee     
where $\lambda$ is the complex dilatino of type IIB string theory and $f^{(12,-12)}$ 
has modular weight
$(12,-12)$~\cite{Green:1997me,Green:1999by}.  
In fact, the higher derivative terms in the effective action which are of the 
form $D^{2k} {\cal{R}}^4$ 
satisfy conjectured non--renormalization theorems, and so it is an interesting problem to determine 
their coupling dependence. Also the terms in the effective 
action which are related to $D^{2k} {\cal{R}}^4$ by 
supersymmetry, for example $\hat{G}^{2k} \lambda^{16}$ ($\hat{G}$ involves 
the three--form field strength 
and certain fermion bilinears), are not renormalized (see~\cite{Sinha:2002zr} for 
the case when $k=2$).       
 
The conjectured non--renormalization theorem for the $D^{2k} {\cal{R}}^4$ term ($k > 0$) in 
the type IIB effective action
predicts that this term does not receive perturbative corrections above $k$ string 
loops~\cite{Berkovits:2006vc,Green:2006gt}. In fact, this has been proven for $0 < k 
< 6$~\cite{Berkovits:2006vc}. 
The structure of these terms has been worked out for some values of $k$. It turns out that
some of these terms 
actually receive even fewer perturbative contributions, as some of the string loop
coefficients vanish. For example at $O (\alpha')$, 
the relevant term in the effective action is given by
\be 
\alpha' \int d^{10} x \sqrt{-g} e^{\phi/2}Z_{5/2} (\tau, \bar\tau) D^4 {\cal{R}}^4, \ee
which receives perturbative contributions only at tree level and at 
two loops~\cite{Green:1999pu,Green:1999pv,D'Hoker:2005ht,Xiao:2005yn} (also 
see~\cite{Berkovits:2005ng,Berkovits:2006bk}). 
At $O (\alpha'^2)$, we have
the term
\be \label{neweqn}
\alpha'^2 \int d^{10} x \sqrt{-g} e^{\phi} {\cal{E}}_{(3/2,3/2)} (\tau, \bar\tau) D^6 
{\cal{R}}^4, \ee
where ${\cal{E}}_{(3/2,3/2)}$ receives contributions from $0,1,2,$ and 3 loops~\cite{Green:2005ba}. 
Similar is the analysis at the next order, where we have the term
\be 
\alpha'^3 \int d^{10} x \sqrt{-g} e^{3\phi/2} Z_{7/2} (\tau, \bar\tau) D^8 
{\cal{R}}^4 \ee
which receives perturbative contributions only at tree level and at 
three loops~\cite{Berkovits:1998ex}. Based on the conjectures stated above, this pattern of 
non--renormalization is believed to persist at higher orders in $\alpha'$ 
(see~\cite{Berkovits:1998ex} for example, for a series of such conjectures).

Note that the coupling dependence of all the terms in the effective action described above
(except \C{neweqn}) is given by the Eisenstein series $Z_s$ for specific values of $s$. It
is easy to see that $Z_s$ satisfies the differential equation
\be 4\tau_2^2 \frac{\p^2}{\p \tau \p \bar\tau} Z_s (\tau, \bar\tau ) = s(s-1) Z_s (\tau, 
\bar\tau).\ee
Thus $Z_s$ is an eigenfunction of the Laplace operator on the fundamental domain of 
$SL(2, \mathbb{Z})$. However, the coefficient of the $D^6 {\cal{R}}^4$ term satisfies the 
differential equation~\cite{Green:2005ba}   
\be 4\tau_2^2 \frac{\p^2}{\p \tau \p \bar\tau} {\cal{E}}_{(3/2,3/2)}  =
12 {\cal{E}}_{(3/2,3/2)} -6 Z_{3/2}^2 .\ee
Thus ${\cal{E}}_{(3/2,3/2)}$ satisfies the Laplace equation in the presence of a source term
on the fundamental domain of $SL(2, \mathbb{Z})$, which can be understood heuristically by 
considering the supersymmetry transformations at higher orders in $\alpha'$.  Clearly
as we go to higher and higher orders in the derivative expansion of the effective action, 
we expect the coupling dependent coefficients to satisfy the Laplace equation in presence of
the source terms~\cite{Green:2005ba,Chalmers:2006aj}. Thus the coupling dependent
coefficients of the $D^{2k} {\cal{R}}^4$ terms for all $k$ generically satisfy Poisson equations 
on the fundamental domain of $SL(2, \mathbb{Z})$ (for low values of $k$, the source terms 
vanish).

Based on heuristic arguments of sypersymmetry and the structure of the three loop four graviton
amplitude of eleven dimensional supergravity compactified on $S^1$ and $T^2$, it seems natural to 
assume that the coefficient of the $D^{10} {\cal{R}}^4$ term in the type IIB effective action 
is given by
\be 
\alpha'^4 \int d^{10} x \sqrt{-g} e^{2\phi} {\cal{E}}_{(3/2,5/2)} (\tau, \bar\tau) D^{10} 
{\cal{R}}^4, \ee  
which satisfies the Poisson equation
\be \label{needeqn}
4\tau_2^2 \frac{\p^2}{\p \tau \p \bar\tau} {\cal{E}}_{(3/2,5/2)}  =
\lambda_1 {\cal{E}}_{(3/2,5/2)} + \lambda_2 Z_{3/2} Z_{5/2}, \ee
on the fundamental domain of $SL(2, \mathbb{Z})$~\cite{Green:2005ba}, where $\lambda_1$ and
$\lambda_2$ are numerical factors. In this paper, we shall solve for $\lambda_1$
and $\lambda_2$ based on known results about the four graviton
scattering amplitude in type IIB superstring theory at tree level and one loop. 
Apart from completely specifying the equation in \C{needeqn}, this will automatically 
lead to predictions for the
two loop and three loop coefficients of the four graviton amplitude in superstring 
perturbation theory, as well as equations that give the contributions due to D--instantons.  

We would like to stress that the ansatz suggested in~\cite{Green:2005ba} for the
coefficient of the $D^{10} {\cal{R}}^4$ term does not give the complete picture, in particular,
it does not satisfy various constraints imposed by unitarity~\cite{Greentalk}. However, 
our analysis does illustrate some of the features of the exact answer. 
 
We also discuss certain generalizations for terms at higher order in 
the $\alpha'$ expansion to understand some features of the exact solution. In particular, we 
consider the $D^{12} {\cal{R}}^4$ and $D^{14} 
{\cal{R}}^4$ terms in the effective action.
We propose that the coupling dependences of these terms also satisy Poisson equation  
on the fundamental domain of $SL(2, \mathbb{Z})$. (For $D^{12} {\cal{R}}^4$, there is
another possibility  where the coupling dependence satisfies the Laplace equation.)    
Simply based on the assumed structure of these equations, we obtain simple vanishing theorems
for the perturbative contributions to these terms.

\section{Some features of the $D^{10} {\cal{R}}^4$ coupling dependence}

In order to determine the differential equation satisfied by ${\cal{E}}_{(3/2,5/2)}$, we 
write it as
\be \label{spliteqn}
{\cal{E}}_{(3/2,5/2)} (\tau, \bar\tau)= {\cal{E}}_{(3/2,5/2)}^{(0)} (\tau_2) + \sum_{k \neq 0} 
{\cal{E}}_{(3/2,5/2)}^{(k)} (\tau_2) e^{2\pi i k \tau_1}.\ee

The ``zero mode'' piece ${\cal{E}}_{(3/2,5/2)}^{(0)}$ is independent of $\tau_1$, and receives 
two kinds of contributions:

(i) the perturbative string loop contributions which involve power law behavior in $\tau_2$,

(ii) the non--perturbative contributions due to D--instantons and anti--D--instantons carrying equal 
and opposite charges. Thus in these terms which receive contributions from double instantons, 
the $e^{2\pi i k \tau_1}$ factor from the D--instanton
of charge $k$ cancels the $e^{-2\pi i k \tau_1}$ factor from the anti--D--instanton
of charge $-k$. So at weak coupling, the leading behavior of this part of the zero--mode 
should be given by
\be \label{leadcont}
\sum_{k \neq 0} f_k \tau_2^{w_k} e^{-4\pi \vert k \vert \tau_2},\ee
and thus these contributions are exponentially suppressed. 

The ``non--zero mode'' part of ${\cal{E}}_{(3/2,5/2)}$ which contains the entire $\tau_1$
dependence, receives contributions from ${\cal{E}}_{(3/2,5/2)}^{(k)}$ for all non--zero values of $k$. 
From the form of \C{spliteqn}, we see that ${\cal{E}}_{(3/2,5/2)}^{(k)}$ gives the non--perturbative 
contribution from the charge $k$ sector. In fact, this can arise from two sources: the charge $k$
single D--instanton contribution, or the double D--instanton contribution 
from two D--instantons of charges $k_1$ and $k_2$ such that $k =k_1 + k_2 \neq 0$.

We first determine the two numerical constants $\lambda_1$ and $\lambda_2$ in \C{needeqn}
which completely specify the differential equation that ${\cal{E}}_{(3/2,5/2)}$ satisfies.
Note that we define the $D^{10} {\cal{R}}^4$ term in the type IIB effective action to be given by the 
specific structure of index contractions such that it leads to a contribution proportional 
to 
\be \label{term5}
\s_2 \s_3 \equiv (\frac{\alpha'}{4})^5 (s^2 + t^2 + u^2) (s^3 + t^3 + u^3)\ee 
in string amplitudes, where $s,t,$ and $u$ are the Mandelstam variables satisfying $s+ t+ u=0$.  
In order to obtain $\lambda_1$ and $\lambda_2$, we shall make 
use of two pieces of information about the four graviton scattering amplitude in type II superstring 
theory; namely, the coefficients of \C{term5} at tree level and 
at one loop in superstring perturbation theory\footnote{At one loop, the four graviton 
scattering amplitude is the same in type IIA and type IIB string theories.}.

The relevant term at tree--level is given by~\cite{Green:1981ya,D'Hoker:1988ta}

\be \label{valtree}
A^{\rm{tree}} = \kappa_{10}^2 e^{-2\phi} \hat{K} \Big ( \ldots + \frac{2}{3} \zeta (3) 
\zeta (5) \s_2 \s_3 +\ldots \Big),\ee
where $\hat{K}$ is the linearized approximation to 
${\cal{R}}^4$~\cite{Green:1981yb,Green:1981ya}. Also, 
the relevant term at one--loop is given by~\cite{Green:1999pv,Green:2006gt}
\be \label{valoneloop}
A^{\rm{one-loop}} = 4 \zeta(2) \kappa_{10}^2 \hat{K} \Big ( \ldots + \frac{29 \zeta(5)}{960} 
\cdot \frac{5 \cdot 4^5}{6 \cdot 5!}  \s_2 \s_3 +\ldots \Big).\ee

\subsection{The perturbative contribution to $D^{10} {\cal{R}}^4$}

The perturbative part of the zero--mode piece ${\cal{E}}_{(3/2,5/2)}^{(0)} (\tau_2)$ 
can receive contributions upto five string loops~\cite{Berkovits:2006vc,Green:2006gt}. Thus using
\C{valtree} and \C{valoneloop} we have that
\be \label{valcons}
{\cal{E}}_{(3/2,5/2)}^{(0)} (\tau_2) = \frac{2}{3} \zeta (3) \zeta (5) \tau_2^4 
+\frac{29 \cdot 5 \cdot 4^6 \zeta (2) \zeta(5)}{960 \cdot 6 \cdot 5!} \tau_2^2 + A
+ \frac{B}{\tau_2^2} + \frac{C}{\tau_2^4} + \frac{D}{\tau_2^6}
+ \ldots,\ee
where the $\ldots$ denotes the non--perturbative terms involving the instanton--anti--instanton 
contributions discussed above. In \C{valcons}, $A,B,C,$ and $D$ are
the coefficients at $2,3,4,$ and 5 string loops respectively. The basic idea is to now use 
\C{needeqn} to write down the differential equation satisfied by the perturbative piece of
${\cal{E}}_{(3/2,5/2)}^{(0)}$. In order to do so, we shall need the expression for $Z_s$ given by
\bea \label{defZ}
Z_s (\tau, \bar\tau) &=& 2\zeta (2s) \tau_2^s + 2\sqrt{\pi} \tau_2^{1-s} \frac{\Gamma 
(s-1/2) \zeta (2s-1)}{\Gamma (s)} \non \\&&+ \frac{4 \pi^s \sqrt{\tau_2}}{\Gamma (s)} \sum_{k \neq 0} 
\vert k \vert^{s-1/2} \mu (k,s) K_{1/2 -s} (2\pi \vert k \vert \tau_2) e^{2\pi ik\tau_1},\eea
where 
\be \mu (k,s) = \sum_{m \vert k} \frac{1}{m^{2s-1}}.\ee 

Note the perturbative contributions to $Z_s$ are given by the first two terms in \C{defZ}. Plugging
in the perturbative contributions from the various terms into \C{needeqn}, and equating 
the coefficients of different powers of $\tau_2$, we get the system of equations
\bea \label{syseqn}
8 -\frac{2}{3} \lambda_1 &=& 4\lambda_2, \non \\
\lambda_2 &=& \frac{29 (2-\lambda_1)}{270} , \non \\
\lambda_1 A &=& -\frac{16}{3} \lambda_2 \zeta (3) \zeta (4), \non \\
(6- \lambda_1) B &=& \frac{32}{3} \lambda_2 \zeta (2) \zeta (4), \non \\
(20 -\lambda_1) C &=& (42 -\lambda_1) D= 0 . 
\eea  
The solution to \C{syseqn} is given by
\be \label{solval}
\lambda_1 = \frac{241}{8}, \quad \lambda_2= -\frac{145}{48}, \quad A= \frac{8 \cdot 
145}{9 \cdot 241} \zeta (3) \zeta (4), \quad B = \frac{16 \cdot 145}{9 \cdot 193} \zeta (2)
\zeta (4), \quad C = D= 0.\ee

Thus we see that ${\cal{E}}_{(3/2,5/2)}$ satisfies the Poisson equation
\be \label{fineqn}
4\tau_2^2 \frac{\p^2}{\p \tau \p \bar\tau} {\cal{E}}_{(3/2,5/2)}  =
\frac{241}{8} {\cal{E}}_{(3/2,5/2)} -\frac{145}{48}  Z_{3/2} Z_{5/2} \ee
on the fundamental domain of $SL (2, \mathbb{Z})$. 

Also from \C{solval}, we see that the four and five--loop coefficients vanish, and $A$ and 
$B$ give predictions for the two loop and three loop amplitudes respectively. Thus the
$D^{10} {\cal{R}}^4$ term in the type IIB effective action receives perturbative contributions 
only upto three loops. 

\subsection{The non--perturbative contribution to $D^{10} {\cal{R}}^4$}

We now consider the non--perturbative part of ${\cal{E}}_{(3/2,5/2)}$, which
receives contributions both from the zero--mode as well as the non--zero mode terms in
\C{spliteqn}. Let us call 
$\tilde{\cal{E}}_{(3/2,5/2)}^{(0)}$ the non--perturbative part of ${\cal{E}}_{(3/2,5/2)}^{(0)}$. 
Then using \C{defZ} we see that $\tilde{\cal{E}}_{(3/2,5/2)}^{(0)}$ satisfies the 
differential equation

\be \label{eqn1}
\Big( \tau_2^2 \frac{\p^2}{\p \tau_2^2} -\frac{241}{8} \Big) 
\tilde{\cal{E}}_{(3/2,5/2)}^{(0)} = -\frac{16 \cdot 145 \pi^3 \tau_2}{3} \sum_{k \neq 0}
\vert k \vert^3 \mu (k, 3/2) \mu (k, 5/2) K_{-1} (2\pi \vert k \vert \tau_2) 
K_{-2} (2\pi \vert k \vert \tau_2).\ee
 
The term on the right hand side of \C{eqn1} contains the total contribution from  
instanton--anti--instanton configurations which carry total charge zero.
Similarly it is easy to see that ${\cal{E}}_{(3/2,5/2)}^{(k)}$ satisfies the 
differential equation

\bea \label{eqn2}
\Big( \tau_2^2 \frac{\p^2}{\p \tau_2^2} -4\pi^2 k^2 -\frac{241}{8} \Big)  
{\cal{E}}_{(3/2,5/2)}^{(k)} = -290 \pi \Big( \frac{2\pi}{3} \{ \zeta (3) \tau_2^2
+ 2 \zeta (2) \} k^2 \mu (k, 5/2) K_{-2} (2\pi \vert k \vert \tau_2) \non \\
+ \{ \zeta (5) \tau_2^3 + \frac{4}{3} \zeta (4) \tau_2^{-1} \} \vert k \vert
\mu (k, 3/2) K_{-1} (2\pi \vert k \vert \tau_2) \non \\
+ \frac{8\pi^2 \tau_2}{3} \sum_{k_1 \neq 0, k_2 \neq 0, k_1 + k_2 =k} \vert k_1 \vert  k_2^2
\mu (k_1, 3/2) \mu (k_2, 5/2) K_{-1} (2\pi \vert k_1 \vert \tau_2)
K_{-2} (2\pi \vert k_2 \vert \tau_2)\Big). \eea

The first two terms on the right hand side of \C{eqn2} give the contributions due to
single instantons of charge $k$, while the last term gives the double instanton contributions 
with total charge $k_1 + k_2 =k$. One can obtain the D--instanton contributions to 
${\cal{E}}_{(3/2,5/2)}$ from the differential equations above.  

For example, using the asymptotic expansion 
\be K_s (z) \sim \sqrt{\frac{\pi}{2z}} e^{-z}\ee 
for large $z$, the leading contribution to 
$\tilde{\cal{E}}_{(3/2,5/2)}^{(0)}$ at weak coupling is given by

\be \tilde{\cal{E}}_{(3/2,5/2)}^{(0)} (\tau_2) \approx -\frac{145 \pi}{12 \tau_2^2} 
\sum_{k \neq 0} \mu (k, 3/2) \mu (k, 5/2) e^{-4\pi \vert k \vert \tau_2},\ee
which is of the form \C{leadcont}.

Thus we see that the coupling dependence of the $D^{10} {\cal{R}}^4$ term in the 
effective action of type IIB superstring theory is given by the
Poisson equation \C{fineqn}, which leads to 
predictions for the two loop and three loop scattering amplitude of four 
gravitons in type IIB superstring theory\footnote{At two loops, the amplitude is the 
same in type IIB and type IIA string theories. It has been shown~\cite{Berkovits:2006vc}
that the perturbative contributions are the same to all loop orders for 
$D^{2k} {\cal{R}}^4$ for $k \leq 4$.}. 
It should be possible to verify the prediction for the two loop amplitude along the lines 
of~\cite{D'Hoker:2005ht,Xiao:2005yn}. 

\section{Some further generalizations}

In order to illustrate some features of the coupling dependence of the $D^{2k} {\cal{R}}^4$ terms in the
type IIB effective action for higher values of $k$ using the method described above, one has to 
know the four graviton amplitude in type IIB superstring theory to sufficiently high order
in the genus expansion, which is a difficult problem. However, as we now show, it is 
easy to obtain certain vanishing theorems for the perturbative contributions which we illustrate 
below with two examples.  
 
\subsection{The $D^{12} {\cal{R}}^4$ term}

The $D^{12} {\cal{R}}^4$ term arises at $O(\alpha'^5)$ in the effective action of type IIB 
string theory. From the tree level amplitude, one can easily see that there are two independent
ways of contracting the various indices, which lead to contributions proportional to $\s_2^3$
and $\s_3^2$. In fact the tree level contributions are 
proportional to $\zeta (9)$ and $\zeta (3)^3$. Thus one linear combination of $\s_2^3$
and $\s_3^2$ should lead to the term in the effective action (where in $D^{12} {\cal{R}}^4$ the 
indices are to be contracted appropriately)~\cite{Berkovits:1998ex}   

\be 
\alpha'^5 \int d^{10} x \sqrt{-g} e^{5\phi/2} Z_{9/2} (\tau, \bar\tau) D^{12} 
{\cal{R}}^4, \ee
which gives $\zeta (9)$ at tree level. So there are only two perturbative contributions 
at tree level and at four loops. Another linear combination of $\s_2^3$ and
$\s_3^2$ yields

\be 
\alpha'^5 \int d^{10} x \sqrt{-g} e^{5\phi/2} {\cal{E}}_{(3/2,3/2,3/2)} 
(\tau, \bar\tau) D^{12} {\cal{R}}^4, \ee
which must give $\zeta (3)^3$ at tree level. This suggests that 
${\cal{E}}_{(3/2,3/2,3/2)}$ should satisfy Poisson equation of the form
\be \label{eqnhigh}
4\tau_2^2 \frac{\p^2}{\p \tau \p \bar\tau} {\cal{E}}_{(3/2,3/2,3/2)}  =
\lambda_1 {\cal{E}}_{(3/2,3/2,3/2)} +\lambda_2 {\cal{E}}_{(3/2,3/2)} Z_{3/2} +\lambda_3 
Z_{3/2}^3.\ee 

One should be able to fix the three undetermined coefficients in \C{eqnhigh} if the
four point graviton amplitude is known upto two loop level at this order in the 
derivative expansion. However, simply based on the structure of \C{eqnhigh} 
without any additional information, we can obtain a 
constraint on the perturbative contributions as we now explain. Assuming
that $D^{12} {\cal{R}}^4$ can receive perturbative contributions only upto 
six loops, we can write the perturbative part of
${\cal{E}}_{(3/2,3/2,3/2)}$ as
\be {\cal{E}}_{(3/2,3/2,3/2)}^{\rm{pert}} (\tau_2) = \cdots + \frac{A}{\tau_2^{11/2}} 
+ \frac{B}{\tau_2^{15/2}},\ee
where $A$ and $B$ are the five and six loop contributions respectively, and the $\cdots$
stands for the other lower loop perturbative contributions. Then \C{eqnhigh} implies that
\be \Big( \frac{11 \cdot 13}{4} - \lambda_1 \Big) A = \Big( \frac{15 \cdot 
17}{4}-\lambda_1 \Big) B =0,\ee
and so $A$ and $B$ cannot be both non--vanishing. Thus at least one of the two
highest loop contributions to ${\cal{E}}_{(3/2,3/2,3/2)}$ must vanish.

\subsection{The $D^{14} {\cal{R}}^4$ term} 

Proceeding exactly along the same lines as before, we note that the $\s_2^2 \s_3$ contribution
to the tree level amplitude is proportional to $\zeta (5)^2$ as well as $\zeta (3) \zeta (7)$.
Thus the $D^{14} {\cal{R}}^4$ term in the effective action is given by
\be 
\alpha'^6 \int d^{10} x \sqrt{-g} e^{3\phi} {\cal{E}}_{(3/2,5/2,7/2)} 
(\tau, \bar\tau) D^{12} {\cal{R}}^4, \ee
where ${\cal{E}}_{(3/2,5/2,7/2)}$ should satisfy the Poisson equation
\be \label{eqnhigher}
4\tau_2^2 \frac{\p^2}{\p \tau \p \bar\tau} {\cal{E}}_{(3/2,5/2,7/2)}  =
\lambda_1 {\cal{E}}_{(3/2,5/2,7/2)} +\lambda_2 Z_{3/2} Z_{7/2} +\lambda_3 
Z_{5/2}^2.\ee 

Again assuming that $D^{14} {\cal{R}}^4$ can receive perturbative contributions only upto 
seven loops, we see that
\be {\cal{E}}_{(3/2,5/2,7/2)}^{\rm{pert}} (\tau_2) = \cdots + \frac{A}{\tau_2^5} 
+ \frac{B}{\tau_2^7} +  \frac{C}{\tau_2^9},\ee
where $A,B$ and $C$ are the five, six and seven loop contributions respectively, and the $\cdots$
stands for the lower loop perturbative contributions. Then \C{eqnhigher} implies that
\be \Big( 30 - \lambda_1 \Big) A = \Big( 56 -\lambda_1 \Big) B = 
\Big( 90 -\lambda_1 \Big) C=0 ,\ee
and so at least two of $A,B,$ and $C$ must vanish. 

Clearly this kind of analysis can be carried out for the $D^{2k} {\cal{R}}^4$ terms
for higher values of 
$k$. This will give vanishing theorems for the perturbative contributions at high orders in the
string loop expansion. It would be interesting to use such constraints alongwith the explicit
coefficients of the four graviton scattering amplitude at low string loops to try to constrain the 
coupling dependence of the various protected higher derivative terms in type IIB superstring 
theory. However, as mentioned before, this kind of analysis does not give the complete
structure of the coupling dependence satisfied by the coefficients of the higher derivative terms, 
although it does illustrate some of the general features.

\section*{Acknowledgements}

I am grateful to Michael B. Green for very useful comments.
The work of A.~B. is supported by NSF Grant No.~PHY-0503584 and the William D. Loughlin membership.





\providecommand{\href}[2]{#2}\begingroup\raggedright\endgroup

\end{document}